\documentclass[twocolumn,showpacs,preprintnumbers,amsmath,amssymb,pra,superscriptaddress]{revtex4}
\usepackage{graphicx}% Include figure files
\usepackage{dcolumn}% Align table columns on decimal point
\usepackage[tight]{subfigure}
\usepackage{amsmath}
\usepackage{verbatim}
\usepackage[usenames,dvipsnames]{color}
\usepackage{bm} % bold greek math
\usepackage{bbm}
\usepackage{floatrow}
\usepackage{bibunits}

\newcommand{\Figure}[2]{
  \begin{figure}[ht]
    \includegraphics[width=1.0\linewidth]{#1}
    \caption{#2}
  \end{figure}
}
\newcommand{\Bigfigure}[2]{
  \begin{figure*}[ht]
    \includegraphics[width=1.0\linewidth]{#1}
    \caption{#2}
  \end{figure*}
}

\newcommand{\wideeq}[1]{
\begin{widetext}
    #1
\end{widetext}}

\newcommand{\Fig}[1]{Fig.~\ref{#1}}

\newcommand{\Eq}[1]{Eq.~(\ref{#1})}
\newcommand{\eq}[1]{(\ref{#1})}
\newcommand{\Sec}[1]{Sec.~\ref{#1}}

\newcommand{\new}[1]{\textcolor{blue}{#1}}

\newcommand{\cut}[1]{\textcolor{red}{cut: #1}}
\newcommand{\todo}[1]{ \textbf{\textcolor{Bittersweet}{TODO: #1}} }
\newcommand{\hide}[1]{ \textbf{\textcolor{Gray}{#1}} }

\newcommand{\nocomma}{}
\newcommand{\tmem}[1]{{\em #1\/}}
\newcommand{\tmop}[1]{\ensuremath{\operatorname{#1}}}

\newcommand{\nobracket}{}

\renewcommand{\new}[1]{{#1}}
\renewcommand{\cut}[1]{{}}
\renewcommand{\todo}[1]{}
\renewcommand{\hide}[1]{}

\begin{document}
\begin{bibunit}[apsrev]
\title{Two-dimensional platform for networks of Majorana bound states}
\author{Michael Hell}
\affiliation{Center for Quantum Devices and Station Q Copenhagen, Niels Bohr Institute, University of Copenhagen, DK-2100 Copenhagen, Denmark}
\affiliation{Division of Solid State Physics and NanoLund, Lund University, Box.~118, S-22100, Lund, Sweden}
\author{Martin Leijnse}
\affiliation{Center for Quantum Devices and Station Q Copenhagen, Niels Bohr Institute, University of Copenhagen, DK-2100 Copenhagen, Denmark}
\affiliation{Division of Solid State Physics and NanoLund, Lund University, Box.~118, S-22100, Lund, Sweden}
\author{Karsten Flensberg}
\affiliation{Center for Quantum Devices and Station Q Copenhagen, Niels Bohr Institute, University of Copenhagen, DK-2100 Copenhagen, Denmark}
\date{\today}

\begin{abstract}
  We model theoretically a two-dimensional electron gas (2DEG) covered by a superconductor and demonstrate that topological superconducting channels are formed when stripes of the superconducting layer are removed. As a consequence, Majorana bound states (MBS) are created at the ends of the stripes. We calculate the topological invariant and energy gap of a single stripe, using realistic values for an InAs 2DEG proximitized by an epitaxial Al layer. We show that the topological gap is enhanced \new{when the structure is made asymmetric. This can be achieved by either imposing a phase difference (by driving a supercurrent or using a magnetic-flux loop) over the strip or by replacing one superconductor by a metallic gate. Both strategies also enable control} over the MBS splitting, thereby facilitating braiding and readout schemes based on controlled fusion of MBS. Finally, we outline how a network of Majorana stripes can be designed.
\end{abstract}

 \maketitle

Majorana bound states (MBS) are states localized at the edges of topological superconductors {\cite{AliceaReview,FlensbergReview,BeenakkerReview,TewariReview}}. They have nonlocal properties that may be utilized for storage and manipulation of quantum information in a topologically protected way {\cite{BravyiKitaev,BravyiKitaev2,Freedman03}}. However, the realization of MBS requires superconducting p-wave pairing, which appears only in exotic materials. Therefore, there is currently a search for ways to engineer p-wave pairing by combining s-wave superconductors with strong spin-orbit materials. Recent experiments looked for
evidence of MBS in, for example, semiconducting nanowires {\cite{mourik12,das12,finck12,Rokhinson,deng12,Albrecht2016,Zhang16}}, topological insulators \cite{Williams}, and magnetic atom chains {\cite{Nadj-Perge,Ruby15}}. These systems may also allow demonstration experiments of the nonlocal properties of MBS, for example using recent suggestions for controlling MBS in prototypical architectures \cite{SauBraiding,Flensberg,BeenakkerBraiding,Li16,Aasen15,Hell16}. However, to go beyond basic demonstration experiments a scalable and flexible platform for large-scale MBS networks is needed.

Here, we suggest one such flexible platform based on a two-dimensional electrons gas (2DEG) with strong spin-orbit coupling in proximity to a superconductor \cite{Shabani16}. Such structures, reviewed in Ref.~\cite{SchaepersBook}, have been realized by contacting InAs surface inversion layers \cite{Chrestin97,Chrestin99} or InAs/InGaAs heterostructures \cite{Takayanagi95,Bauch05} with superconducting Nb or Al.
Recently, it has become possible to grow an Al top layer epitaxially \cite{Shabani16}, forming a clean interface with the 2DEG. The proximitized 2DEG (denoted by pS) develops a hard superconducting gap as revealed by
experiments on pS-N quantum point contacts \cite{Kjaergaard16} or gateable pS-N-pS junctions \cite{Shabani16} with clear signatures of multiple Andreev reflection \cite{KjaergaardMAR} \new{and nontrivial Fraunhofer patterns \cite{Suominen16}.} The transport properties of these structures have been studied extensively \cite{SchaepersBook} but their potential as a MBS platform has not.

\Figure{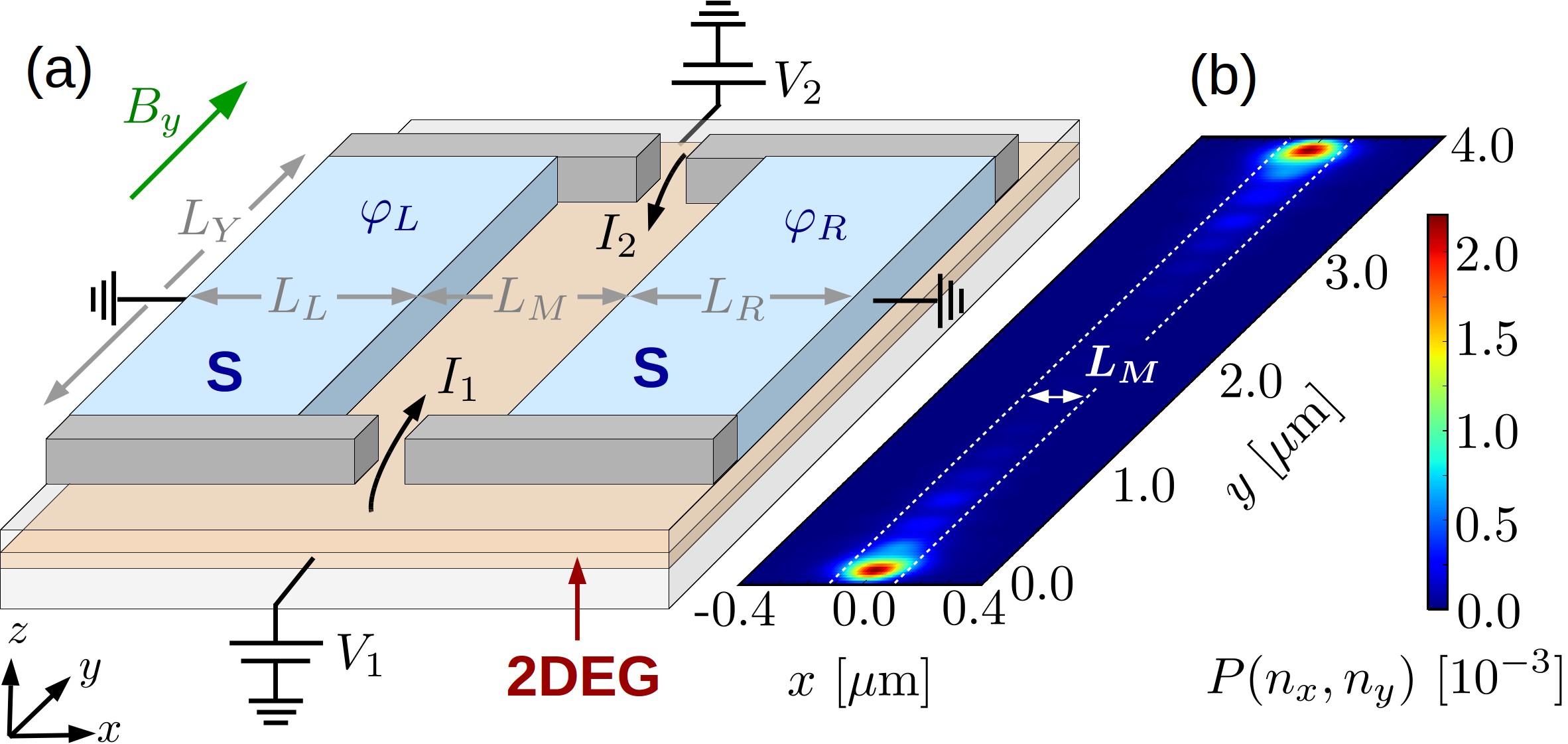}{Illustration of a stripe hosting MBS in a 2DEG-based platform. Panel (a) shows the device with the superconducting layer removed along a stripe, and point contacts formed at the ends to facilitate tunneling spectroscopy of the MBS shown in (b), where the probability density \cite{Hell162DEGsup} $P(n_x,n_y)$ of the MBS wave function is plotted. The calculation is done on a lattice with $N_x = 260$ and $N_y = 160$ sites an the parameters are $E_Z =200~\mu$eV, $\Gamma_L = \Gamma_R = 180~\mu$eV, $\alpha = 1.42 \cdot 10^{- 4} c$, $\mu = 0$, $L_L = L_R = 1~\mu$m, $L_M = 250~$nm, $L_Y = 4~\mu$m, $m^{\ast} = 0.023 m_e$, and $\varphi_L = \varphi_R = 0$.\label{fig:model}}

We show how to design and control MBS in a pS system with a stripe of the superconducting layer removed to form an effective one-dimensional topological superconductor. This forms a pS-N-pS junction as sketched in Fig. \ref{fig:model}(a), which can be fabricated by standard lithographic techniques. We show, similar to other semiconductor-based setups {\cite{Sau,Alicea,1DwiresLutchyn,1DwiresOreg,Sedlmayr15}}, that this system undergoes several topological phase transitions  when increasing a magnetic field parallel to the stripe for parameters readily available in the lab. We base our findings on a numerical tight-binding calculation of the energy spectrum, the topological invariant, as well as transport calculations \cite{Hell162DEGsup}.  We discuss how the topological energy gap depends on various parameters, which is vital for the topological protection and the manipulation time scales of the MBS. We show that slight modifications of the structure shown in \Fig{fig:model} give a large enhancement of the topological gap and, moreover, grant electrical control over the phase-transition point. The key is to break the inversion symmetry perpendicular to the stripe direction, and we study two methods to accomplish this: \textit{(i)} a phase bias (generated by a supercurrent) across the stripe, or \textit{(ii)} replacement of one of the superconducting top layers by a gate electrode. Both methods allow to fuse the MBS electrically, which can be used to manipulate MBS in 2DEG structures. In the last part of the paper, we discuss designs of more advanced MBS networks for fusion-rule testing and braiding.

{\tmem{Model.}} We model the (unproximitized) 2DEG by a single electron band
with effective mass $m^{\ast}$ and electro-chemical potential $\mu$. The device
has a finite extension with $- ( L_L + L_M / 2) \leq x \leq ( L_R + L_M / 2)$ and $|
y | \leq L_Y / 2$, where it is described by the Bogoliubov-de Gennes
Hamiltonian ($e = \hbar = k_B = c = 1$):
\begin{eqnarray}
  H ( x, y) & = & \left[ - \tfrac{1}{2 m^{\ast}}  \left(\partial_x^2 + \partial_y^2 \right) -
  \mu \right] \tau_z \nonumber\\
  &  & - i \alpha ( \sigma_x \partial_y - \sigma_y \partial_x) \tau_z + E_Z
  \sigma_y/2 .  \label{eq:ham}
\end{eqnarray}
In the second line, we add the Rashba spin-orbit coupling (with velocity
$\alpha$) and the Zeeman energy ($E_Z$) due to a magnetic field along the
stripe. The Hamiltonian acts on the four-component spinor ${\mathbf{\psi}} = [ \psi_{e, \uparrow}, \psi_{e, \downarrow}, \psi_{h,
\downarrow}, - \psi_{h, \uparrow}]^T$ containing the electron
$( e)$ and hole $( h)$ components for spin $\sigma = \uparrow, \downarrow$.
The Pauli matrices $\tau_i$ and $\sigma_i$ ($i = x, y, z$) act on
particle-hole and spin space, respectively.

We include the proximity effect of the superconducting top layer within the
Green's function formalism. Integrating out the superconductor in the
wide-band limit, the Green's function of the 2DEG is given by
{\cite{SauProximityEffect,StanescuProximityEffect,Hansen16}}
\begin{eqnarray}
  G_R ( x, \omega) & = & \left[ \omega - H ( x, y) - \Sigma ( x, \omega) + i
  0_+ \right]^{-1},  \label{eq:gr}
\end{eqnarray}
with self energy
\begin{eqnarray}
  \Sigma (x, \omega) & = & \Gamma ( x) \frac{\Delta [ \cos \varphi ( x) \tau_x \new{-}
  \sin \varphi ( x) \tau_y] - \omega}{\sqrt{\Delta^2 - ( \omega + i 0)^2}} .
\end{eqnarray}
The self energy is zero in the stripe region, $\Gamma ( | x | < L_M / 2) = 0$,
and nonzero under the two superconducting layers coupled to the 2DEG with
symmetric tunneling rates $\Gamma ( | x | > L_M / 2) = \Gamma$.
\new{In this way, we only make an assumption about the superconducting order parameter in the metallic top layer, while the superconducting pairing in the 2DEG is determined by \Eq{eq:gr}.
The two top layers are assumed to} have the same gap $\Delta$ but \new{possibly} different phases $\varphi ( x < - L_M
/ 2) = \varphi_L$ and $\varphi ( x > L_M / 2) = \varphi_R$. Such a phase bias
can be realized experimentally by running a supercurrent across the stripe.
Our self energy does not include a proximity-induced shift of the
chemical potential under the superconductor. This is motivated by recent experiments \cite{KjaergaardMAR}
showing that pS-N-pS junctions have a high transparency, which indicates
a rather small mismatch in Fermi velocities.

{\tmem{Symmetry class and topological invariants.}} We first investigate the general topological properties of the stripe. Since its aspect ratio is large ($L_Y \gg L_M$), the system is quasi-1D, similar to coupled {\cite{Wakatsuki14,Kotetes15}} or multiband {\cite{StanescuProximityEffect,Lutchyn11Multiband}} nanowires. The topological properties are in general determined by the zero-frequency Green's function $G^R ( x, 0)$ {\cite{Wang13,Budich13}}. The self energy $\Sigma ( x, 0)$ takes in this limit the form of a non-dissipative pairing term and the topological properties are thus determined by {\cite{Wang14}}
\begin{eqnarray}
  H_{\tmop{eff}} &=& H ( x, y) +\Gamma ( x) [ \cos \varphi ( x) \tau_x \new{-} \sin \varphi ( x) \tau_y]. \label{eq:heff}
\end{eqnarray}

This effective Hamiltonian respects particle-hole symmetry since it anticommutes with the antiunitary operator $P = \sigma_y \tau_y \mathcal{K}$ ($\mathcal{K}$ denotes the complex conjugation). If no generalized time-reversal symmetry is present, the system is thus in symmetry class D ($P^2=1$) with a $\mathbbm{Z}_2$ topological invariant $W_{\mathbbm{Z}_2}$ {\cite{Altland97,Ryu10}}.

However, our system can also be in the higher-symmetry class BDI with an integer topological invariant $W_{\mathbbm{Z}}$. This is the case if the system \new{has spatial symmetry in $x$-direction ($L_L = L_R$)}, assuming here no disorder in the $x$-direction. The effective Hamiltonian then possesses an additional generalized 'time-reversal' symmetry: it commutes with the antiunitary operator \new{$T = \sigma_z I_x \mathcal{K}$} ($I_x$ is the reflection in $x$-direction) {\footnote{The time-reversal and particle-hole conjugation operator are not unique since $H_{\tmop{eff}}$ has an additional unitary symmetry $U = I_y\sigma_y$, where $I_y$ is the reflection operator in $y$ direction.}}. \new{This symmetry holds even in the presence of a phase bias \cite{Pientka16}.}
Unlike the physical time reversal, the generalized operator $T$ squares to identity, $T^2 = 1$. If both $P$- and $T$-symmetry are present, also chiral symmetry is present, i.e., $H_{\tmop{eff}}$ anticommutes with \new{$C = -i P T = \sigma_x \tau_y I_x$}.

\Figure{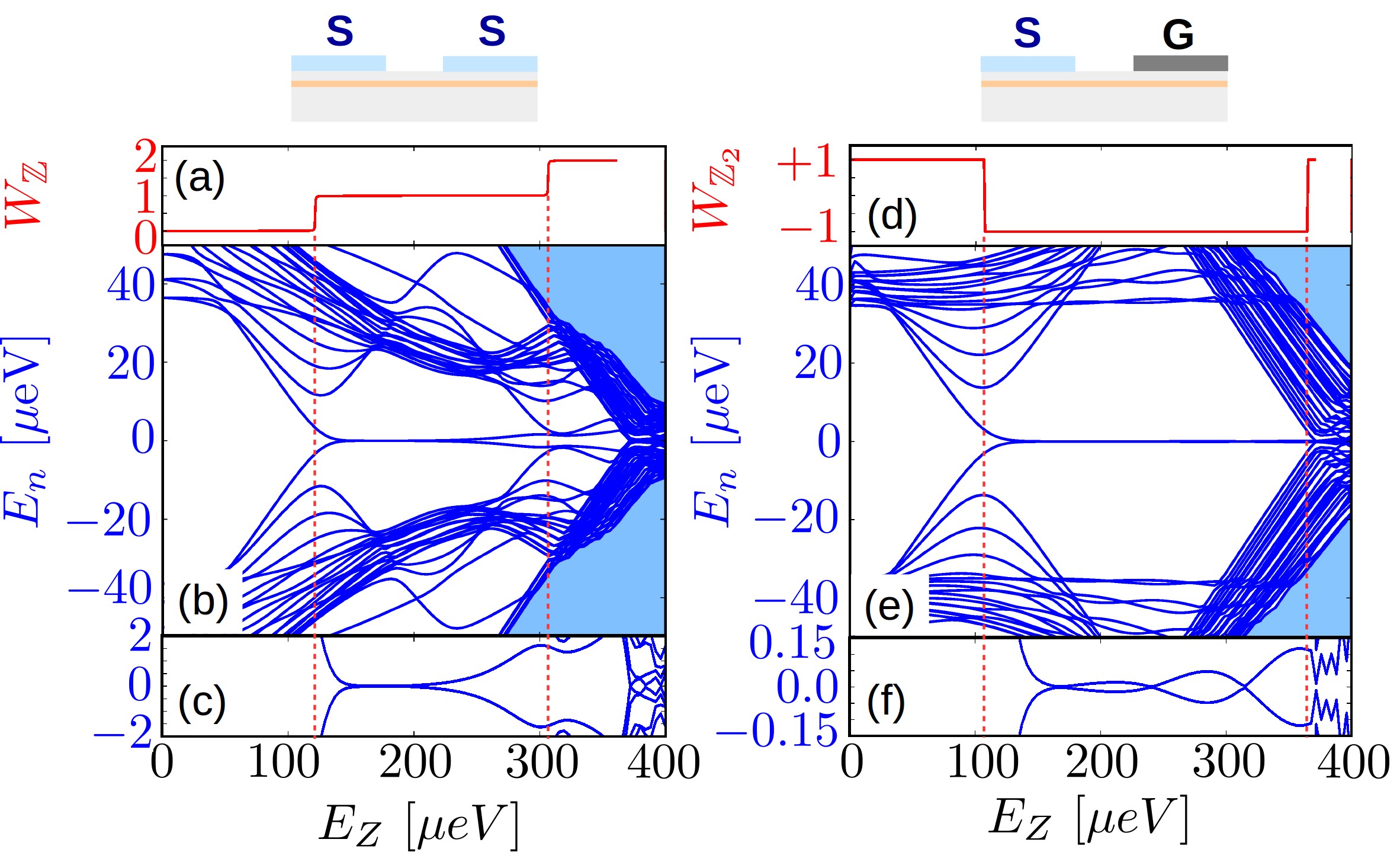}{{Topological phase transition with increasing Zeeman field $E_Z$}. The top pictograms illustrate the two cases studied here: on the left a symmetric device ($L_L = L_R$, class BDI) and on the right an asymmetric top-gated device ($L_R = 0$, class D). The upper panels show the topological invariant $W_{\mathbbm{Z}}$ in (a) and $W_{\mathbbm{Z}_2}$ in (d).
  The mid panels (b) and (e) depict the 50 lowest eigenenergies of $H_{\tmop{eff}} (x, y)$ [\Eq{eq:heff}]. Higher excited states form a quasicontinuum in the light-blue shaded areas.
  Closeups of the midgap-mode energies are shown in the lower panels (c) and (f).
  All parameters are as in Fig. \ref{fig:model}(b).
  \label{fig:spectrum}}

\Bigfigure{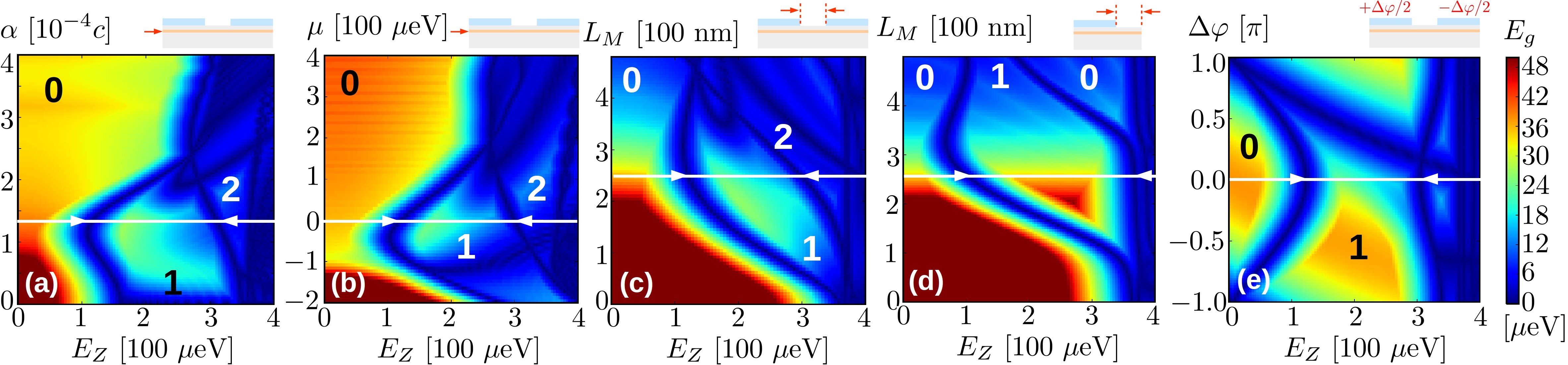}{Enhancing the topological energy gap. We define the topological energy gap by  $E_g \equiv \new{\min_{n,k_y}}  | E_n ( k_y) |$, where $E_n ( k_y)$ are the eigenenergies of $H_{\tmop{eff}} ( x, -i \partial_y \rightarrow k_y)$ [see \Eq{eq:heff}]. The gap is shown as a function of the Zeeman energy $E_Z$ and (a) the spin-orbit velocity $\alpha$, (b) the electro-chemical potential $\mu$, (c) and (d) the width of the stripe $L_M$ for a symmetric and asymmetric device, respectively, and (e) the phase difference $\varphi_L = - \varphi_R = \Delta \varphi /2$. We use $N_x=300$ lattice points in (a),(b), and (e), but a finer resolution of $N_x=3/2 (L_L+L_M+L_R)$[nm] for (c) and (d). If not varied, all parameters are as in Fig.~\ref{fig:model}(b). The color scale is cut off at 50 $\mu\text{eV}$ to enhance the contrast inside the regimes with a single MBS (we inserted the number of MBS, marked phase-separation lines with white arrows, and indicate the parameters used in all other plots by white lines). \label{fig:top_gap}}

To predict a topological phase transition,
we compute topological invariants $W_{\mathbbm{Z}}$ (BDI) and
$W_{\mathbbm{Z}_2}$ (D). To obtain $W_{\mathbbm{Z}}$, we follow
Ref.~{\cite{Tewari12TopInv}}:
Because the chirality operator satisfies $C^{\dag} C = C^2 = 1$, its only eigenvales are
$\pm 1$ and in these two subblocks the Hamiltonian is
off-diagonal (since $[ C, H_{\tmop{eff}}]_+ = 0$):
\begin{eqnarray}
  C \text{ \ = \ } \left(\begin{array}{cc}
    1 & 0\\
    0 & - 1
  \end{array}\right), &  & H_{\tmop{eff}} \text{ \ = \ }
  \left(\begin{array}{cc}
    0 & A\\
    A^{\dag} & 0
  \end{array}\right) .
\end{eqnarray}
The $\mathbbm{Z}$ invariant follows from the winding number of the phase
$\theta (k_y)$ of the determinant of $A$, $\det A ( k_y) / | \det A ( k_y) | =
e^{i \theta ( k_y)}$ as $W_{\mathbbm{Z}}  =  \int_0^{\infty} d k_y  d {\theta} ( k_y)/ d k_y / \pi$.
The integer $W_{\mathbbm{Z}}$ characterizes the number of MBS which appear at
the boundaries of a long stripe with two topologically trivial regions.

When $T$-symmetry is broken, an even number of MBS on the same boundary couple to each other,
turning them into finite-energy modes. This leaves either zero or one MBS,
characterized by the $\mathbbm{Z}_2$ invariant $W_{\mathbbm{Z}_2}$. We compute
$W_{\mathbbm{Z}_2}$ in the standard way {\cite{1DwiresKitaev}} by
representing $H_{\tmop{eff}}$ as a matrix $M ( k_y)$ in Majorana
representation. The topological invariant is given by the relative sign of the
Pfaffian of $M (k_y)$ at the $T$-invariant points $k_y = 0$ and $k_y =
\infty$ \cite{Hell162DEGsup}.

{\tmem{Topological phase transition.}} Figure \ref{fig:spectrum} demonstrates that the stripe region undergoes a topological phase transition for material parameters in the range of recent experiments {\cite{Shabani16,Kjaergaard16,KjaergaardMAR}}.  We obtain our results numerically using a tight-binding approximation of the effective Hamiltonian \eq{eq:heff} {\cite{Ferry97Book,Hell162DEGsup}}.

We first discuss the $T$-symmetric case (left panels of Fig. \ref{fig:spectrum}). Starting from the topologically trivial regime ($W_{\mathbbm{Z}} = 0$), one can identify a first phase transition at Zeeman energy $E_{Z, \tmop{cr}} \sim 120~\mu$eV [$W_{\mathbbm{Z}} = 0 \rightarrow 1$, Fig.~\ref{fig:spectrum}(a)]. {For a $g$-factor of about 10 {\cite{Kjaergaard16}}, this requires a magnetic field of $\sim$200 mT, much lower than the critical fields in Al thin films {\cite{Tedrow81}}.} At first sight it may be surprising that the critical Zeeman energy $E_{Z, \tmop{cr}}$ is smaller than the induced superconducting gap $\Gamma$ under the superconducting top layers. The reason is that the Andreev bound states in the stripe experience the pairing potential only where they penetrate into the proximitized region and, as a consequence, the effective gap is smaller than $\Gamma$.

In agreement with the change in $W_{\mathbbm{Z}}$, a pair of states comes close to zero energy around $E_Z = E_{Z, \tmop{cr}}$ [Fig. \ref{fig:spectrum}(b)]. The wave functions of these states are localized at the ends of the stripe [Fig. \ref{fig:model}(b)] and, because of the finite length of the stripe, their energies oscillate around zero [Fig. \ref{fig:spectrum}(c)]. From  Fig. \ref{fig:spectrum}(b), we extract an energy gap to excited states -- the topological energy gap -- of about 20$~\mu$eV, which corresponds to about 200~mK.

A second phase transition takes place for $E_Z \sim 320~\mu$eV
[$W_{\mathbbm{Z}} = 1 \rightarrow 2$,  Fig.~\ref{fig:spectrum}(a)], where, for a finite stripe length, a second pair of states approaches zero
energy [Fig.~\ref{fig:spectrum}(b)]. However, for the
parameters chosen in  Fig. \ref{fig:spectrum}, the states
remain split in energy because the $W_{\mathbbm{Z}} = 2$ regime is close to the
breakdown of the induced superconducting gap (at $E_Z = 2 \Gamma \new{\sim} 360 ~ \mu
$eV). By increasing the stripe width, one can reach the regime of two
MBS for lower $E_Z$ \cite{Hell162DEGsup}.

To confirm the presence of the MBS, one can probe the conductance of the stripe by two quantum point contacts [see \Fig{fig:model}(a)]. We compute the transport spectrum numerically \cite{Hell162DEGsup} and find a zero-bias peak in the local conductance with a peak value up to $2 e^2 / h$ \cite{ZeroBiasAnomaly1,Nilsson08,Tewari08}
while the nonlocal response is strongly suppressed for uncoupled MBS, different from extended Andreev bound states.

For future applications for topological quantum-information processing, it is desirable to enhance the topological energy gap as much as possible. This can be achieved, for example, with an asymmetric device structure as depicted above  Fig.~\ref{fig:spectrum}(d): One replaces one of the superconducting layers by a top gate that creates a potential barrier for the electrons in the 2DEG underneath. We model this here by terminating the system at the right end of the stripe setting $L_R = 0$. The Hamiltonian is now in class D since $T$-symmetry is broken. The $\mathbbm{Z}_2$ invariant $W_{\mathbbm{Z}_2}$ indicates a phase transition around $E_Z \sim 120~\mu$eV [$W_{\mathbbm{Z}_2} = + 1 \rightarrow - 1$,  Fig.~\ref{fig:spectrum}(d)] and a single pair of states approaches zero energy [Fig.~\ref{fig:spectrum}(e)]. Compared with the symmetric device, their energy splitting is much smaller [$\sim 0.1~\mu$eV,  Fig.~\ref{fig:spectrum}(f)] and the topological gap is increased [$\sim 40~\mu$eV,  Fig.~\ref{fig:spectrum}(e)].
\emph{An asymmetric device design thus seems promising for stabilizing MBS}.

\Figure{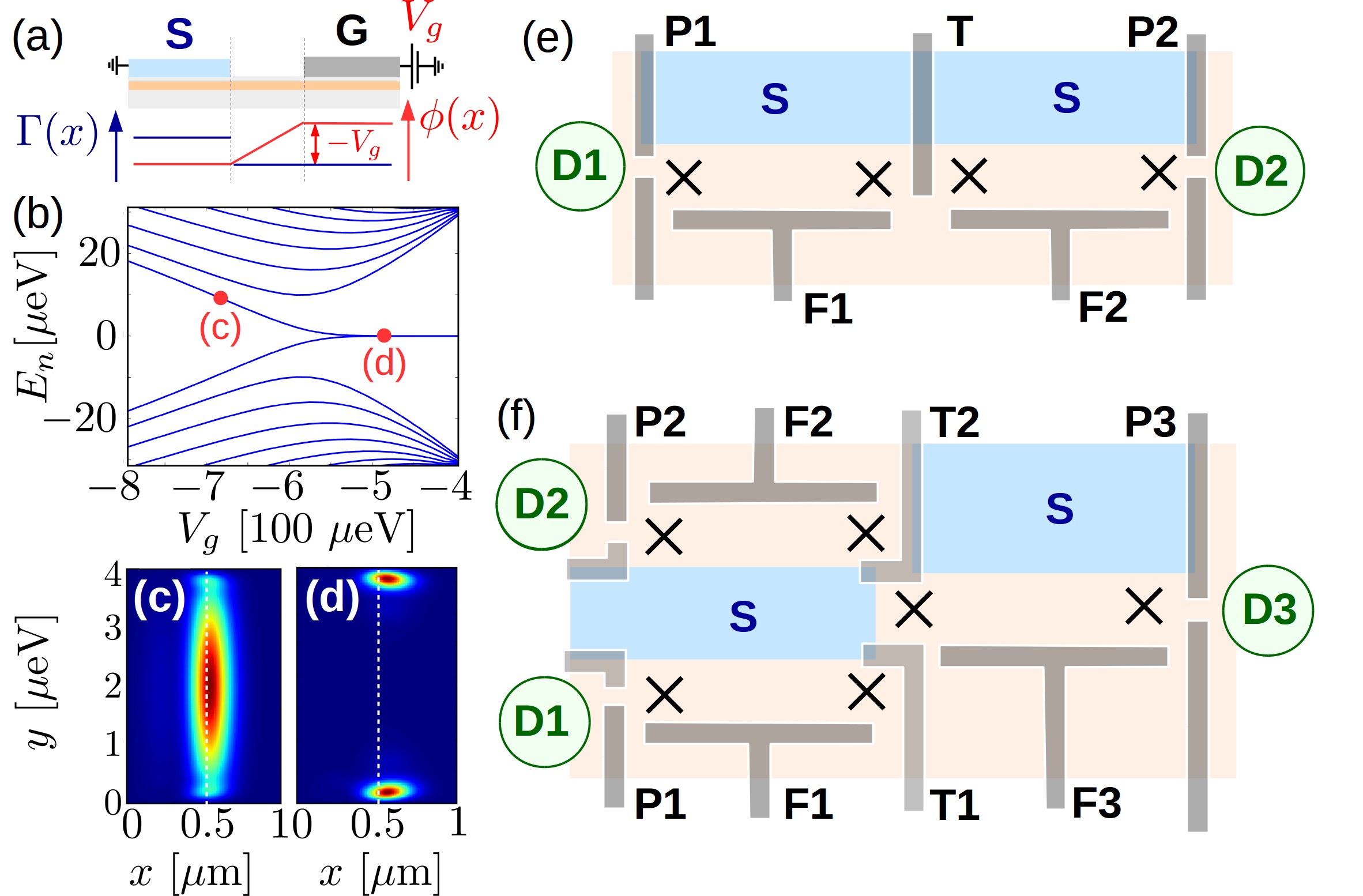}{Electrically controlled manipulation of MBS in 2DEG devices. (a) Sketch of a MBS stripe whose width is controlled by a linear voltage drop tuned by a top gate (G) with (b) the corresponding low-energy spectrum. Below, we show the probability density $P(n_x,n_y)$ for the state closest to zero energy when the MBS are (c) coupled and (d) uncoupled. We use \new{$N_x=N_y=150$} lattice points, all other parameters are as in \Fig{fig:model}(b). (e) By segmenting a MBS stripe into two parts using a gate (T), one could implement a qubit consisting of four MBS, which are denoted by crosses. The intra-island coupling of the MBS could be tuned by fusion gates (F1) and (F2) and the inter-island coupling by the tunneling gate (T). For readout, the two gates (P1) and (P2) allow for coupling to two detectors (D1) and (D2). (f) Braiding setup with two gates (T1) and (T2) controlling the tunnel coupling between the three center MBS.\label{fig:structures}}

{\tmem{Phase diagrams and topological gap.}} To study the optimal conditions for observing MBS in experiments more systematically, we next investigate the topological energy gap (see caption of \Fig{fig:top_gap}). Both the gap and the boundaries of the topological phases given by the zero-gap condition exhibit a nontrivial dependence on the different parameters [Fig.~\ref{fig:top_gap}]. Due to the finite-frequency and finite-size effects, the actual topological gap may be smaller, but by comparing with transport calculations (including both effects) we find only a small reduction \cite{Hell162DEGsup}.

We first focus on the case \new{with inversion symmetry} [Figs.~\ref{fig:top_gap}(a)--(c)]. The phase-transition point depends on the spin-orbit coupling $\alpha$ [Fig.~\ref{fig:top_gap}(a)]; \new{however, a change in $\alpha$ can be compensated by a shift of the electro-chemical potential $\mu \rightarrow \mu + m^* \alpha^2 / 2 $ \cite{Pientka16}. By contrast, the topological gap depends in a nontrivial way on $\alpha$:
It} becomes maximal around $\alpha \approx 1.2 \cdot 10^{- 4} c$ and is is strongly suppressed for $\alpha > 2 \cdot 10^{- 4} c$ for the experimentally relevant parameters used in \Fig{fig:top_gap}(a) \new{for fixed $\mu$}. Fortunately, the spin-orbit velocities extracted from current experiments {\cite{Shabani16}},
$\alpha \approx 1.42 \cdot 10^{- 4} c$, are close to being optimal. From Fig.~\ref{fig:top_gap}(b) we see that it is crucial to tune the electro-chemical potential near zero \new{if no phase bias is applied}, which requires a strong gate coupling, similar to the situation for nanowires \cite{1DwiresLutchyn,1DwiresOreg}. Finally, we find that a stripe width $L_M$ of around 200 nm is optimal for a large topological gap [Fig.~\ref{fig:top_gap}(c)].

As mentioned before, breaking the generalized $T$-symmetry can further increase the topological gap. \new{For example, the} asymmetric, gated device [Fig.~\ref{fig:top_gap}(d)] exhibits a topological gap that can be larger by a
factor of up to 2.
\new{However, the gap can also be manipulated without breaking the generalized $T$-symmetry in a phase-biased device [Fig.~\ref{fig:top_gap}(e)].} Depending on the sign of the phase bias $\Delta \varphi$, the direction of the supercurrent is reversed, which through the spin-orbit term $i \alpha \sigma_y \partial_x$ affects the spectrum differently for a nonzero Zeeman term. \new{This can both increase and decrease the gap, which} could be used to determine the sign of $\alpha$ experimentally.
Second, both for gating and phase bias the regime of MBS can be
reached for smaller Zeeman energies [see $L_M \approx 300 \tmop{nm}$ in
Fig.~\ref{fig:top_gap}(d) and $\Delta \varphi \approx - \pi$ in
Fig.~\ref{fig:top_gap}(e)]. Finally, the phase transition
point can be moved {\tmem{in-situ}}, either by changing $L_M$ through the gate
or by a supercurrent controlling $\Delta \varphi$.

{\emph{Electrical control for MBS networks.} The above-mentioned sensitivity of the phase-transition point suggests the possibility to control the coupling of the MBS electrically. To illustrate this, we assume that the nearby top gate creates a triangular confining potential for the MBS [Fig.~\ref{fig:structures}(a) and (b)]. By increasing the potential drop across the stripe by lowering $V_g$, the MBS can be tuned from localized boundary modes [Fig.~\ref{fig:structures}(d)] at zero energy into modes at finite energy and delocalized along the stripe [Fig. \ref{fig:structures}(c)]. Controlling the MBS coupling with a gate can be used for the initialization and readout of the MBS also in larger networks of MBS stripes. Readout requires a way to detect the fermion parity of the MBS, which can be achieved by charge detection \cite{Ben-Shach14}, possibly using an auxiliary quantum dot \cite{Hoving16readout}. }

{To implement a qubit in a 2DEG structure, one can segment the MBS stripe into two parts using a finger gate [Fig. \ref{fig:structures}(e)]. Tuning this gate controls the coupling of the MBS in the middle (indicated by crosses), which could be used to carry out fusion-rule and coherence-test experiments \cite{Aasen15,Hell16}. Finally, in order to realize braiding of MBS similar to Refs. \cite{SauBraiding,BeenakkerBraiding,Aasen15,Hell16}, one needs to couple three MBS. Our numerical simulations indicate \cite{Hell162DEGsup} that the topological phase is stable against a rotation of the magnetic field of about 10$^\circ$ away from the stripe direction. We therefore suggest using a 'tuning-fork' design [Fig. \ref{fig:structures}(f)] instead of a T-junction structure usually considered for braiding-type experiments. This keeps the stripes in parallel.}

\tmem{Conclusion and outlook.} We have shown that 2DEG structures with
strong spin-orbit coupling, proximity-induced superconductivity, and magnetic
fields provide an alternative platform hosting MBS that is readily available in the lab.
A phase bias or a gate electrode can be used to couple the MBS in the stripe by electrical means. Together with the technological advantage of flexible top-down fabrication of 2DEG structures, this might open the door for larger MBS networks.
  \new{Note: After the first version of this preprint appeared, a preprint appeared that treats a device similar to the one considered here \cite{Pientka16}.}

\begin{acknowledgments}
{We acknowledge stimulating discussions with M. Kjaergaard, P. Kotetes, C. M. Marcus, F. Nichele, and H. J. Suominen, and support from the Crafoord Foundation (M. L. and M. H.), the Swedish Research Council (M. L.), and The Danish National Research Foundation. \new{Computational resources were partly provided by the Swedish National Infrastructure
for Computing (SNIC) through Lunarc, the Center for Scientific and Technical Computing at Lund University.} }
\end{acknowledgments}

\putbib[cite]
\end{bibunit}

%\bibliography{cite}
%\bibliographystyle{apsrev}

%%%% SUPPLEMENTAL MATERIAL %%%%%%%%%%%%

\clearpage
\begin{bibunit}[apsrev]
  \setcounter{equation}{0}
  \setcounter{figure}{0}
  \setcounter{section}{0}
\renewcommand{\thefigure}{S\arabic{figure}}
\renewcommand{\theequation}{S\arabic{equation}}

\title{Two-dimensional platform for networks of Majorana bound states: \\ Supplemental Material }
\author{Michael Hell}
\affiliation{Center for Quantum Devices and Station Q Copenhagen, Niels Bohr Institute, University of Copenhagen, DK-2100 Copenhagen, Denmark}
\affiliation{Division of Solid State Physics and NanoLund, Lund University, Box.~118, S-22100, Lund, Sweden}
\author{Martin Leijnse}
\affiliation{Center for Quantum Devices and Station Q Copenhagen, Niels Bohr Institute, University of Copenhagen, DK-2100 Copenhagen, Denmark}
\affiliation{Division of Solid State Physics and NanoLund, Lund University, Box.~118, S-22100, Lund, Sweden}
\author{Karsten Flensberg}
\affiliation{Center for Quantum Devices and Station Q Copenhagen, Niels Bohr Institute, University of Copenhagen, DK-2100 Copenhagen, Denmark}
\date{\today}
  
\maketitle

\section{Numerical approach}

In this Section, we explain our numerical procedure to compute the energy
spectrum, topological invariants, and topological gaps presented in the main
part. Our approach is based on a tight-binding
approximation of the effective Hamiltonian [Eq. (4) in the main part], which
we introduce first. Since diagonalizing the 2D tight-binding Hamiltonian is
computationally demanding, we speed up our calculations by solving the
eigenvalue problem in two steps: We first solve the one-dimensional tight-binding problem
transverse to the stripe direction, project on a number of low-energy modes,
and in the last step solve a second one-dimensional tight-binding problem along the stripe
direction. We finally discuss how the different symmetry classes for symmetric
and asymmetric devices manifest in the energy spectrum when a second phase
transition appears.

\subsection{Two-dimensional tight-binding model}\label{sec:2dtb}

The tight-binding model for our calculations is obtained by discretizing the
continuous spatial coordinates $( x, y)$ as a lattice with $N_x$ points in the
$x$ direction (perpendicular to the stripe) and $N_y$ points in the $y$
direction (along the stripe). The lattice points are given by $( x_n = n 
d_x, y_m = m d_y)$ with $n = 1, \ldots, N_x$, $m = 1, \ldots, N_y$ and
lattice constants $d_i = N_i / L_i$. Due to the discretization, we have to
replace the derivatives in the Hamiltonian by finite differences:
\begin{eqnarray}
  \frac{\partial \psi}{\partial x} ( x_n) & \approx & \frac{\psi ( x_{n + 1})
  - \psi ( x_{n - 1})}{2 d_x}, \\
  \frac{\partial^2 \psi}{\partial x^2} ( x_n) & \approx & \frac{\psi ( x_{n +
  1}) + \psi ( x_{n - 1}) - 2 \psi ( x_n)}{d^2_x} . 
\end{eqnarray}
Analogous formulas apply for the $y$ derivative. Using this procedure, we can rewrite the
inverse retarded Green's function [Eq. (2) in the main part] in
tight-binding approximation. Using the hopping amplitudes
\begin{eqnarray}
  t_i \text{ \ = \ } \frac{1}{2 m^{\ast} d_i^2}, &  & t_{\tmop{SOC}, i} =
  \frac{\alpha}{2 d_i}, \nocomma \text{ \ \ } ( i = x, y), 
\end{eqnarray}
it reads
\wideeq{
%  \begin{eqnarray}
    \begin{align}
  G_R^{- 1} ( \omega) & = & \omega - H - \Sigma ( \omega) & &
  \nonumber\\
  & = & - \textstyle{\sum}_{n_x n_y \tau \sigma} & \{ [ \tau ( 2 t_x + 2 t_y -
  \mu) - Z^{- 1} ( n_x, \omega) \omega + i 0] | n_x n_y \sigma \tau \rangle
  \langle n_x n_y \sigma \tau | \nobracket \nonumber\\
  & &  & + i \sigma E_Z/2 | n_x n_y \tau \bar{\sigma} \rangle \langle n_x n_y
  \tau \sigma | + \delta ( n_x, \omega) | n_x n_y \bar{\tau} \sigma \rangle
  \langle n_x n_y \tau \sigma | \nonumber\\
  & &  & - [ \tau t_x | n^+_x n_y \tau \sigma \rangle \langle n_x n_y \tau
  \sigma | + \text{H.c.}] - [ \tau \sigma t_{\tmop{SOC}, x} | n_x^+ n_y \tau
  \bar{\sigma} \rangle \langle n_x n_y \tau \sigma | + \text{H.c.}]
  \nonumber\\
  & &  & - [ \tau t_y \nobracket | n_x n^+_y \tau \sigma \rangle \langle n_x
  n_y \tau \sigma | + \text{H.c.}] - [ i \tau t_{\tmop{SOC}, y} | n_x n^+_y
  \tau \bar{\sigma} \rangle \langle n_x n_y \tau \sigma | + \text{H.c.}] \} . 
  \label{eq:grtb}
  \end{align}
%\end{eqnarray}
}
Here, $| n_x n_y \tau \sigma \rangle$ denotes a state with an electron ($\tau=+$) or hole ($\tau=-$) with spin $\sigma = \pm$ localized at lattice point $(n_x,n_y)$. We introduce the short-hand notations $\bar{\sigma} = - \sigma$, $\bar{\tau} = - \tau$, $n_i^{\pm} = n_i \pm
1$, the Z factor {\cite{SauProximityEffect}}
\begin{eqnarray}
  Z^{- 1} ( n_x, \omega) & = & 1 + \frac{\Gamma ( n_x)}{\sqrt{\Delta^2 - (
  \omega + i 0)^2}}, 
\end{eqnarray}
and express the frequency-dependent superconducting pairing as:
\begin{eqnarray}
  \delta ( n_x, \omega) & = & ( Z^{- 1} ( n_x, \omega) - 1) \nonumber\\
  &  &  \Delta [ \cos \varphi ( n_x) +i \tau \sin \varphi ( n_x) ] . 
  \label{eq:reddelta}
\end{eqnarray}
In the limit $\omega \rightarrow 0$, the Green's function is related to the
effective Hamiltonian [Eq. (4) in the main part] by $G^{- 1}_R ( 0) = i 0 -
H_{\tmop{eff}}$. Both $G^{- 1}_R$ and $H_{\tmop{eff}}$ are thus represented by
$( 4 N_x N_y) \times ( 4 N_x N_y)$ matrices for $N_x N_y$ lattice points
with spin ($\sigma = \pm$) and particle-hole ($\tau = \pm$) degree of freedom.
For inferring the topological properties, we use $H_{\tmop{eff}}$, while we
use the full Green's function later for our transport calculations (see \Sec{sec:transport}). The eigenstates of $H_{\tmop{eff}}$ are represented by a
4-component vector at each lattice site
\begin{align}
  \mathbf{\psi} ( n_x, n_y) & = & \left(\begin{array}{c}
    \psi_{+ +} ( n_x, n_y)\\
    \psi_{+ -} ( n_x, n_y)\\
    \psi_{- +} ( n_x, n_y)\\
    \psi_{- -} ( n_x, n_y)
  \end{array}\right) = \left(\begin{array}{c}
    \psi_{\uparrow} ( n_x, n_y)\\
    \psi_{\downarrow} ( n_x, n_y)\\
    \psi^{\dag}_{\downarrow} ( n_x, n_y)\\
    - \psi^{\dag}_{\uparrow} ( n_x, n_y)
  \end{array}\right)
\end{align}
and the probability densities shown for the eigenstates in Figs. 1(b), 4(c), 4(d) in the main part are
given by
\begin{eqnarray}
  P ( n_x, n_y) & = & \sum_{\sigma \tau} | \psi_{\tau \sigma} ( n_x, n_y) |^2
  .
\end{eqnarray}

\subsection{Numerical diagonalization with low-energy projection}

The matrix for the effective Hamiltonian can, in principle, be diagonalized by
standard numerical procedures. However, the number of lattice points that can
be treated is limited by computational power and we therefore perform the
diagonalization in two steps with an intermediate approximation. For this
purpose, we split the effective Hamiltonian into two parts:
\begin{eqnarray}
  H_{\tmop{eff}} & = & \sum_{n_y} \{ H_{\tmop{eff}, y} | n_y \rangle \langle
  n_y | \nobracket \nonumber\\
  &  & + \nobracket [ T_{\tmop{eff}, y} | n_y + 1 \rangle \langle n_y | +
  \text{H.c.}] \} .  \label{eq:heff}
\end{eqnarray}
The first transverse part $H_{\tmop{eff}, y}$ contains all terms that do not
change $n_y$ [related to the first three lines in Eq. (\ref{eq:grtb})] and the
second part $T_{\tmop{eff}, y}$ contains the hopping terms in the $y$
direction [related to the fourth line in Eq. (\ref{eq:grtb})]. We note that
the effective 1D tight-binding Hamiltonian $H_{\tmop{eff}, y}$ is independent
of $n_y$. We first diagonalize this part as $U_y^{\dag} H_{\tmop{eff}, y} U_y
= D_{\tmop{eff}, y}$ with standard procedures, where $D_{\tmop{eff}, y}$
contains the eigenvalues of $H_{\tmop{eff}, y}$. We next apply the unitary
transformation $U = \sum_{n_y} U_y | n_y \rangle \langle n_y |$ to Hamiltonian
(\ref{eq:heff}), which yields
\begin{eqnarray}
  \tilde{H}_{\tmop{eff}} = U^{\dag} H_{\tmop{eff}} U & = & \sum_{n_y} \{
  D_{\tmop{eff}, y} | n_y \rangle \langle n_y | \nobracket 
  \label{eq:hefftransform}\\
  &  & + \nobracket [ \tilde{T}_{\tmop{eff}, y} | n_y + 1 \rangle \langle n_y
  | + \text{H.c.}] \}, \nonumber
\end{eqnarray}
with hopping matrix $\tilde{T}_{\tmop{eff}, y} = U_y^{\dag} T_{\tmop{eff}, y}
U_y$.

We next apply our approximation: We neglect high-energy eigenstates of
$H_{\tmop{eff}, y}$, i.e., we project $D_{\tmop{eff}, y}$ onto the $4 N_x' < 4
N_x$ eigenvalues closest to zero energy, somewhat similar to Ref.
{\cite{Disorder7}}. We denote the corresponding $4 N_x' \times 4 N_x$
projector in the following by $P$. Applying $P$ reduces the dimension of the
matrices $D_{\tmop{eff}, y}$ and $\tilde{T}_{\tmop{eff}, y}$:
\begin{eqnarray}
  P \tilde{H}_{\tmop{eff}} P & = & \sum_{n_y} \{ D^P_{\tmop{eff}, y} | n_y
  \rangle \langle n_y | \nobracket \\
  &  & + \nobracket [ \tilde{T}^P_{\tmop{eff}, y} | n_y + 1 \rangle \langle
  n_y | + \text{H.c.}] \}, \nonumber
\end{eqnarray}
with $D^P_{\tmop{eff}, y} = P D_{\tmop{eff}, y} P$ and
$\tilde{T}^P_{\tmop{eff}, y} = P \tilde{T}_{\tmop{eff}, y} P$ , both $4 N_x'
\times 4 N_x'$ matrices. We include into the projection not only bound states
confined in the stripe but also a part of the spectrum above the gap. This is
necessary because the hopping matrix $\tilde{T}^P_{\tmop{eff}, y}$ couples
subgap states of $H_{\tmop{eff}, x}$ to states above the (induced)
superconducting gap. Moreover, we find that low-energy spectrum of the full
effective $H_{\tmop{eff}}$ cannot be reproduced in a satisfactory way without
these states above the gap. For all plots we chose $N_x' = 40$
except for Fig. 4 where we chose $N_x' = 30$. Diagonalizing Eq.
(\ref{eq:hefftransform}) yields then the energy spectra shown in Fig. 2 in the
main part. We note that the results for the bound states of the stripe do not depend on the widths $L_L$ and $L_R$ of the proximitized regions as long as the wave functions have decayed at the boundary of the simulated area.

\subsection{Topological invariants}

We next explain how to compute the topological invariants $W_{\mathbbm{Z}_2}$
(symmetry class D) and $W_{\mathbbm{Z}}$ (symmetry class BDI), which we
discussed in the main part. This calculation does not require an additional
projection as above since the topological properties are inferred from
a corresponding one-dimensional bulk system. This bulk system is obtained by
extending the stripe to infinity in the positive and negative $y$ direction. We
can thus replace the derivative $- i \partial_y$ by the wave vector $k_y$ and
obtain the following effective Hamiltonian in tight-binding representation in
the $x$ direction:
\begin{eqnarray}
  H_{\tmop{eff}} ( k_y) & = & \sum_{n_x n_y \tau \sigma}  \label{eq:heff1d}\\
  &  & \left\{ \left[ \tau \left( 2 t_x + \tfrac{k_y^2}{2 m^{\ast}} - \mu
  \right) \right] | n_x \tau \sigma \rangle \langle n_x \tau \sigma | \right.
  \nonumber\\
  &  & + ( i \sigma E_Z + \alpha k_y) | n_x \tau \bar{\sigma} \rangle \langle
  n_x \tau \sigma | \nonumber\\
  &  & + \delta ( n_x) | n_x \bar{\tau} \sigma \rangle \langle n_x \tau
  \sigma | \nonumber\\
  &  & - [ \tau t_x | n^+_x \tau \sigma \rangle \langle n_x \tau \sigma | +
  \text{H.c.}] \nonumber\\
  &  & - [ \tau \sigma t_{\tmop{SOC}, x} \nobracket | n^+_x \tau \bar{\sigma}
  \rangle \langle n_x \tau \sigma | + \text{H.c.}] \} \nonumber
\end{eqnarray}
The computation of $W_{\mathbbm{Z}}$ follows along the lines of Ref.~{\cite{Tewari12TopInv}} as explained in the main part. Since the Hamiltonian
is represented by a finite-dimensional matrix, we can also compute the winding
phase $e^{i \theta ( k_y)} = \det A ( k_y) / | \tmop{det}A ( k_y) |$ below Eq.
(5) in the main part by computing the determinant of a finite-dimensional
matrix.

When $T$-symmetry is broken, an even number of MBS couple to each other,
turning them into finite-energy modes. This leaves over zero or one MBS,
characterized by the $\mathbbm{Z}_2$ invariant $W_{\mathbbm{Z}_2}$. We compute
$W_{\mathbbm{Z}_2}$ in the standard way {\cite{1DwiresKitaev}} by first
representing $H_{\tmop{eff}}$ as a matrix $M ( k_y)$ in Majorana
representation:
\begin{eqnarray}
  H_{\tmop{eff}} ( k_y) & = & \frac{i}{2}  \sum_{\eta \eta' \sigma \sigma'
  n_x n_x'} M_{n_x \eta \sigma, n_x' \eta' \sigma'} | n_x \eta \sigma \rangle
  \langle n_x' \eta' \sigma' |, \nonumber \\
  & &
\end{eqnarray}
Here, we introduced new states (suppressing the $n_x$ index, which remains
unaffected),
\begin{eqnarray}
  \left(\begin{array}{c}
    | \tau = +, \sigma = + \rangle\\
    | \tau = -, \sigma = + \rangle\\
    | \tau = +, \sigma = - \rangle\\
    | \tau = -, \sigma = - \rangle
  \end{array}\right) & = & F^{\ast} \left(\begin{array}{c}
    | \eta = +, \sigma = + \rangle\\
    | \eta = -, \sigma = + \rangle\\
    | \eta = +, \sigma = - \rangle\\
    | \eta = -, \sigma = - \rangle
  \end{array}\right)  \ \ \ \
\end{eqnarray}
with the unitary matrix {\footnote{We define $F \tmop{instead} \tmop{of}
\text{$F^{\ast}$}$ since this simplifies the form of Eq. (\ref{eq:m}). Note
that the matrix $F$ characterizes how the electron {\tmem{annihilators}} $[
\psi_{\uparrow} ( n_x), \psi_{\downarrow} ( n_x), \psi_{\downarrow}^{\dag}  (
n_x), - \psi_{\uparrow}^{\dag} ( n_x)]^T$ can be expressed in terms of
Majorana operators $[ \gamma_{\uparrow, 1} ( n_x), \gamma_{\uparrow, 2} (
n_x), \gamma_{\downarrow, 1} ( n_x), \gamma_{\downarrow, 2} ( n_x)]^T$. The
states, by contrast, are related through the complex-conjugate matrix
analogous to the electron creation operators. \ }}
\begin{eqnarray}
  F & = & \left(\begin{array}{cccc}
    1 & 0 & 0 & - 1\\
    - i & 0 & 0 & - i\\
    0 & 1 & 1 & 0\\
    0 & - i & i & 0
  \end{array}\right) 
\end{eqnarray}
The matrix $M_{n_x, n_x'} ( k_y)$ is related to the Hamiltonian matrix by
\begin{eqnarray}
  M_{n_x n_x'} & = & \frac{2}{i} F  H_{n_x n_x'} F^{\dag}, 
  \label{eq:m}
\end{eqnarray}
and satisfies the relations
\begin{eqnarray}
  M^{\dag} ( k_y) & = & - M ( + k_y),  \label{eq:mhc}\\
  M^T ( k_y) & = & - M ( - k_y) .  \label{eq:mt}
\end{eqnarray}
The first of these relations follows simply from the Hermiticity of the
Hamiltonian, $H_{\tmop{eff}} ( k_y) = H^{\dag}_{\tmop{eff}} ( - k_y)$. The
second relation is a consequence of the the particle-hole symmetry,
$\mathcal{P} H_{\tmop{eff}} ( k_y) \mathcal{P}^{\dag} = - H_{\tmop{eff}} ( -
k_y)$, with the particle-hole conjugation operator $\mathcal{P} = \sigma_y
\tau_y \mathcal{K}$ as introduced in the main part. Equation (\ref{eq:mt}) can
be shown by complex conjugating Eq. (\ref{eq:mhc}) and exploiting the relation
$\mathcal{K} F = F \mathcal{P}$.

The $\mathbbm{Z}_2$ invariant can be next expressed as
{\cite{1DwiresKitaev}}
\begin{eqnarray}
  W_{\mathbbm{Z}_2} & = & \tmop{sgn} \frac{\tmop{Pf} ( M ( k_y =
  0))}{\tmop{Pf} ( M ( k_y \rightarrow \infty))},  \label{eq:wz2}
\end{eqnarray}
where $\tmop{Pf}$ denotes the Pfaffian.
%{\color{red} in our case $k_y \rightarrow \infty$ and not to $k_y = \pi / a$. Our k space is not compact. Is that a problem?}
  Note that for the time-reversal invariant momenta, the matrix
$M$ is real and antisymmetric according to Eqs. (\ref{eq:mt}) and
  (\ref{eq:mhc}). Equation (\ref{eq:wz2}) is thus well-defined. Since the kinetic-energy term dominates for large $k_y$, the Hamiltonian approaches that of quasi-free electrons and the Pfaffian $\tmop{Pf} ( M ( k_y \rightarrow \infty)) =1$ and evaluating Eq. (\ref{eq:wz2}) thus amounts to computing the Pfaffian for $k_y=0$.
  Provided the
Hamiltonian is in the higher-symmetry class BDI, $W_{\mathbbm{Z}_2}$ is still
a topological invariant, which is related to $W_{\mathbbm{Z}}$ by
{\cite{Tewari12TopInv}}
\begin{eqnarray}
  W_{\mathbbm{Z}_2} & = & ( - 1)^{W_{\mathbbm{Z}}} . 
\end{eqnarray}

\subsection{Symmetry class BDI vs. D}

In the main part, we discussed that the symmetric device structure is in
symmetry class BDI and can have an integer number of MBS ($W_{\mathbbm{Z}}$),
while the asymmetric device structure is in symmetry class D and can have zero
or one MBS ($W_{\mathbbm{Z}_2}$). However, in Fig. 2 of the main part, the two
MBS for case BDI are rather ``undeveloped'' since they appear close to the
closing of the superconducting gap. Here we show that the regime of two MBS
can instead be fully developed by increasing the stripe width $L_M$.

\Figure{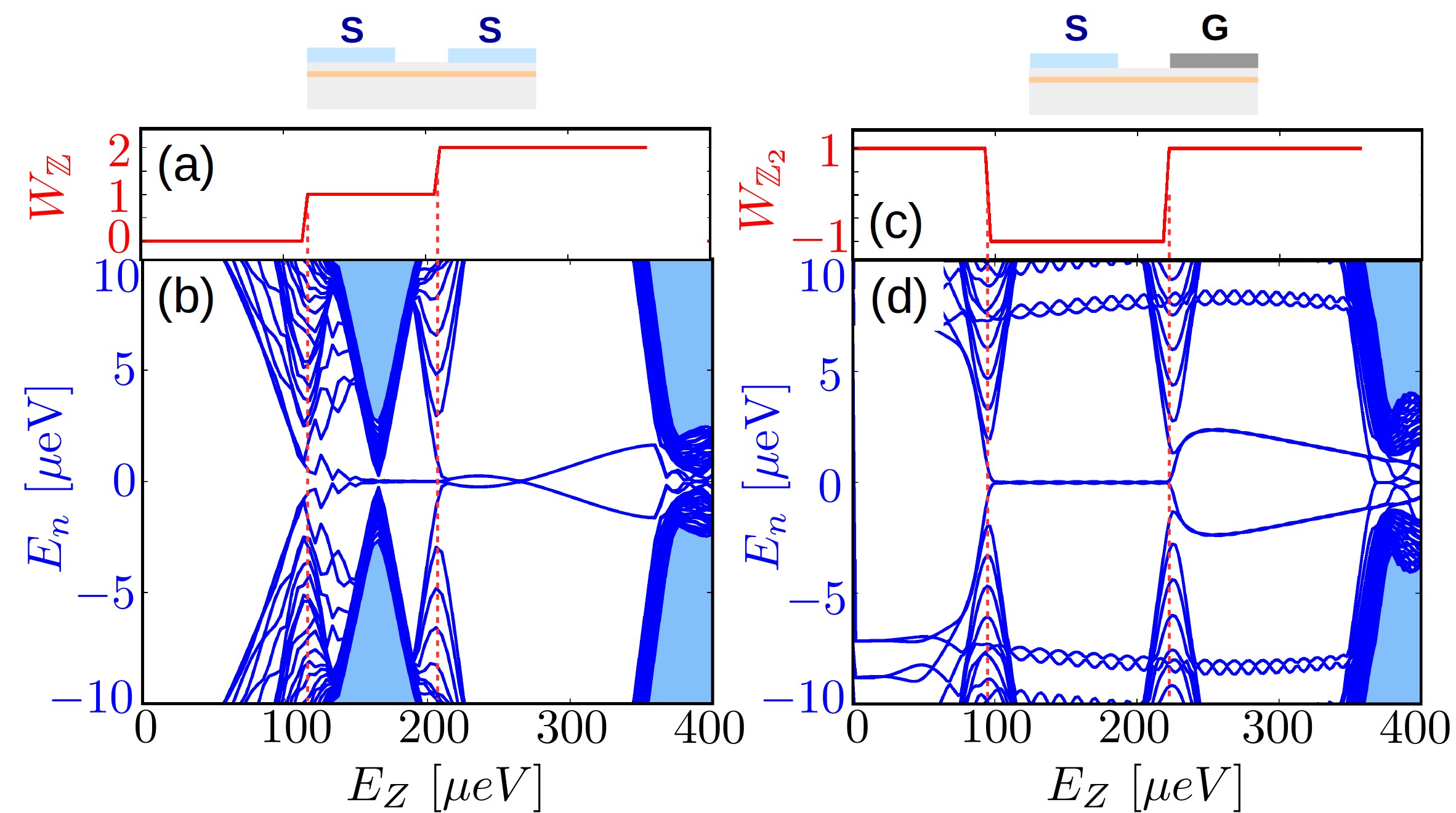}{{Symmetry
  class BDI vs. D.} The two top pictograms illustrate the two cases studied
  here: a symmetric device ($L_L = L_R$, $L_M = 375$nm, class BDI) on the left
  and an asymmetric device ($L_R = 0$, $L_M = 450$nm, class D) on the right.
  The upper panels show the topological invariants computed for $N_x = 300$
  lattice points, in (a) $W_{\mathbbm{Z}}$ for the BDI case and in (b)
  $W_{\mathbbm{Z}_2}$ for the D case. Since the topological invariants lose
  their meaning when the Zeeman energy exceeds the induced superconducting
  gap, we do not show them for $E_Z > 180~\mu$\text{eV}. The panels (b) and
  (d) depict the 50 lowest eigenenergies of the full 2D Hamiltonian
  $H_{\tmop{eff}} ( x, y)$. Higher excited states form a quasi-continuum in
  the light blue shaded areas. We have used a stripe length of $L_Y = 20~\mu$m and widths as mentioned above,
  $N_x = N_y = 150$. All other parameters are as in Fig. 1(b) of the main
  part.\label{fig:spectrum_z}}

Increasing the stripe width is favorable because this lowers the confinement
energy of the second-lowest transverse mode in the Majorana stripe. This mode
evolves into the second pair of MBS. In this way, the phase-transition is
shifted to lower Zeeman energies and a larger topological energy gap can be
achieved. A larger gap reduces the localization length of the MBS wave
functions, which reduces their energy splitting due to finite wave-function
overlap.

Following the above reasoning, we changed the dimensions of the stripe from
250 nm $\times$ 4 $\mu$m for Fig. 2 in the main part to $375$ nm $\times$ 20
$\mu$m \ in Fig. \ref{fig:spectrum_z} shown here. This lowers the critical
Zeeman energy for the second phase transition point, $W_{\mathbbm{Z}} = 1
\rightarrow 2$, to $E_Z \approx 200 \mu$eV \ [Fig. \ref{fig:spectrum_z}(a)].
This is around 100 $\mu$eV lower than for the smaller stripe width [Fig. 2(a) in the
main part]. Around the critical Zeeman field, a second pair of states
approaches zero energy for a symmetric device [Fig. \ref{fig:spectrum_z}(b)].
If instead the $T$-symmetry is broken, the number of MBS alternates between
zero or one as we show here for an asymmetric device [Fig.
\ref{fig:spectrum_z}(d), here taking $450$ nm $\times$ 20 $\mu$m]. The changes
in the energy spectrum coincide with the changes in the $\mathbbm{Z}_2$
invariant $W_{\mathbbm{Z}_2}$ [Fig. \ref{fig:spectrum_z}(c)]. Different from
the $W_{\mathbbm{Z}} = 2$ regime in Fig. \ref{fig:spectrum_z}(b), the states do
not oscillate around zero energy but remain at finite energy instead.

Although the above findings are more of theoretical interest at this point,
they may also have experimentally relevant implications. The regime of a
single MBS is probably more suitable for potential applications for quantum
computation. One would like to avoid the presence of two MBS pairs since this complicates initialization and readout and the MBS
could couple to each other under non-perfect conditions breaking the $T$-symmetry. In this respect, an asymmetric device structure would
avoid such complications and turns out to be favorable. It may also be
interesting to study the transition between the two symmetry classes by simply
changing the stripe width. Hence, a 2DEG platform provides an interesting
experimental playground also from a fundamental physics perspective.

\section{Transport spectra}\label{sec:transport}

The subgap spectrum of the Majorana stripe can be investigated experimentally
most easily by transport measurements. We therefore complement our discussion
in the main part, which focuses on the energy spectrum, by a calculation of
the transport features. The situation we consider is sketched in Fig. 1(a) in
the main part: One could probe the Majorana stripe by connecting it to two
quantum points contacts and measure the response of the currents $I_i$ flowing
from lead $i = 1, 2$ into the stripe as a function of the voltages $V_j$
applied to lead $j = 1, 2$. In short, we find a zero-bias peak in the local
conductance with a peak value up to $2 e^2 / h$, similar to nanowires
{\cite{ZeroBiasAnomaly1,ZeroBiasAnomaly3,ZeroBiasAnomaly4,ZeroBiasAnomaly5,ZeroBiasAnomaly6}}.
By contrast, the nonlocal response is strongly suppressed as long at the two
MBS are uncoupled but appears when the MBS are coupled due to wave-function
overlap, also analogous to nanowires
{\cite{ZeroBiasAnomaly1,Nilsson08,Tewari08,Zocher13,Liu13}}. We first explain
our numerical approach in Sec. \ref{sec:scattering} and discuss our results
for the conductance in Sec. \ref{sec:non-local}.

\subsection{Scattering theory}\label{sec:scattering}

To compute the transport properties, we apply a scattering-matrix formalism
to the setup sketched in Fig. 1 in the main part. The scattering
region is formed by the entire 2DEG part, which is in total connected to four
terminals: two normal leads (terminals 1 and 2) via the quantum point contacts
and two superconductors as top layers on the 2DEG. Here, we are interested
only in the transport through terminals 1 and 2 and compute the conductance
matrix
\begin{eqnarray}
  G_{i j} & = & \partial I_i / \partial V_j \text{ \ } ( i, j = 1, 2) . 
\end{eqnarray}
Due to current conservation in the steady state, the current flowing into the
superconductors is given by $I_1 + I_2$, which is in general nonzero. Since
we assume the superconductors to be grounded, Coulomb-blockade effects are
suppressed in the stripe and a scattering approach may be applied.

We use a simple model for the leads and assume that only one
transverse mode can propagate through each quantum point contact. The two
leads thus provide together eight incoming and outgoing channels, accounting
for electrons and holes ($\tau = \pm$) with spin $\sigma = \uparrow,
\downarrow$. The scattering matrix is then an $8 \times 8$ matrix,
\begin{eqnarray}
  S & = & \left(\begin{array}{cc}
    R_{1 1} & T_{1 2}\\
    T^{\dag}_{1 2} & R_{2 2}
  \end{array}\right), 
\end{eqnarray}
with $4 \times 4$ reflection matrices $R_{i i}$ and $4 \times 4$ transmission
matrices $T_{i j}$. These matrices can be divided into electron- and hole
sectors:
\begin{eqnarray}
  R_{i i} = \left(\begin{array}{cc}
    R_{i i}^{e e} & R^{e h}_{i i}\\
    R^{h e}_{i i} & R_{i i}^{h h}
  \end{array}\right) &,  & T_{i j} = \left(\begin{array}{cc}
    T_{i j}^{e e} & T_{i j}^{e h}\\
    T_{i j}^{h e} & T_{i j}^{h h}
  \end{array}\right) . 
\end{eqnarray}
We compute the scattering matrix by employing the Weidenm\"{u}ller-Mahaux formula
{\cite{Hansen16,Beenakker15rev,Aleiner02rev}}:
\begin{eqnarray}
  S & = & \mathbbm{1} - 2 \pi i W^{\dag} \frac{1}{\omega - G^{- 1}_R + i \pi W
  W^{\dag} + i 0} W.  \label{eq:smat}
\end{eqnarray}
The formula incorporates the the retarded Green's function $G_R$ of the
scattering region, which is a $( 4 N_x N_y) \times ( 4 N_x N_y)$ matrix in our
tight-binding model with $N_x$ and $N_y$ sites in the $x$ and $y$ direction,
respectively (see Sec. \ref{sec:2dtb}). We note that \Eq{eq:smat} includes the full
frequency dependence of the self energy $\Sigma ( \omega)$ contained in the Green's
function $G_R$ and accounts for a finite extension of the system in the $y$
direction. This provides a more accurate estimate of the topological energy
gap than the effective Hamiltonian [Eq. (4) in the main part], which neglects
finite-frequency and finite-size effects.

%Furthermore, in Eq. (\ref{eq:smat}) $\eta$ is an infinitesimally small
%positive real number but we take $\eta$ to be finite in our calculations.
%This leads to a broadening of the conductance peaks in addition to the
%tunneling-induced broadening due to the term $\pi W W^{\dag}$. We will discuss
%this further below in Sec. \ref{sec:non-local}.

The coupling to the leads is described by the $4 N_x N_y \times 8$ matrix $W$,
which contains the couplings of each site of the scattering region to the lead
channels:
\begin{eqnarray}
  W_{i \tau \sigma, n_x n_y \tau' \sigma'} & = & ( \delta_{i, 1} \delta_{n_y,
  1} + \delta_{i, 2} \delta_{n_y, N_y}) \delta_{n_x, \text{stripe}}
  \nonumber\\
  &  & \times \sqrt{\Gamma_p} \tau \delta_{\tau \tau'} \delta_{\sigma \sigma'} 
\end{eqnarray}
The first line expresses that lead $i = 1$ couples to the bottom-most row of
sites $( n_y = 1)$, while lead $i = 2$ couples to the top-most row of sites
($n_y = N_y$). The coupling is restricted in the transverse direction to the sites
inside the Majorana stripe, which we denote in a short-hand way by
$\delta_{n_x, \text{stripe}}$ (in practice, the coupling is also determined by the
width of the quantum-point contact). The Kronecker symbols in the
second line express that the tunneling is spin-conserving and does not mix
particles and holes. Moreover, both leads are coupled to the stripe
symmetrically with the same particle-hole and spin-independent tunneling rate
$\Gamma_p$. Our simple model ignores the effects of spin-orbit coupling and
magnetic field in the leads.

Once the scattering matrix is computed, the differential conductance follows
from the formula {\cite{Lesovik97}}:
\begin{eqnarray}
  G_{i j} & = & \frac{\partial}{\partial V_j} \int d \omega [ f_j ( \omega) -
    f_j ( \omega + e V_j)] G_{s, i j} ( \omega, V_j) . \nonumber \\
  & &
\end{eqnarray}
with the spectral conductance
\begin{eqnarray}
  G_{s, i i} & = & \frac{e^2}{h} \tmop{Tr} [ \mathbbm{1} - R^{e e}_{i i} +
  R^{h e}_{i i}],  \label{eq:gii}\\
  G_{s, i j} & = & \frac{e^2}{h} \tmop{Tr} [ \mathbbm{\text{ \ } } - T^{e
  e}_{i j} + T^{h e}_{i j}] \text{ \ } ( i \neq j),  \label{eq:gij}
\end{eqnarray}
and Fermi function $f_j ( \omega) = 1 / ( 1 + e^{\omega / T_j})$ for lead $j$
at temperature $T_j$ (we assume the reference electro-chemical potential of all leads
to be zero).

In general, the spectral conductance $G_{s, i j} ( \omega, V_j)$ can have an
explicit voltage dependence since the voltages may change the tunnel coupling
or the spectral properties of the scattering region. We ignore this effect
here so that the conductance follows as a folding of the spectral conductance
with the derivative of the Fermi function. For zero temperature $T_j = 0$, as
assumed in the following, the Fermi function becomes a delta function and we
get $G_{i j} = G_{s, i j} ( V_j)$.
%However, as mentioned above, we chose a
%finite value of $\eta$, which leads to a broadening of the peaks and thus has
%qualitatively a similar effect as a finite-temperature broadening.

Using Eqs. (\ref{eq:gii}) and (\ref{eq:gij}), one can decompose the
zero-temperature conductance into contributions related to different transport
processes,
\begin{eqnarray}
  G_{i i}^{\tau \tau'} \text{ \ = \ } \bar{\tau} \frac{e^2}{h} \tmop{Tr}  R^{\tau
  \tau'}_{i i}, &  & G_{i j}^{\tau \tau'} \text{ \ = \ } \bar{\tau} \frac{e^2}{h}
  \tmop{Tr} T^{\tau \tau'}_{i j},  \label{eq:getaetapr}
\end{eqnarray}
where we use $\tau, \tau' = e, h$ as an index but $\tau=\pm$ in mathematical expressions. Since we
use this decomposition for our interpretation of the conductance spectra shown
below in Sec. \ref{sec:non-local}, we briefly review these processes. An
electron incoming from lead 1 has four possibilities as we discuss next. (i)
$G^{e e}_{1 1}$: It can be normally reflected as an electron leading to no
current [the contribution from $G^{e e}_{1 1}$ cancels with the term $\sim \mathbbm{1}$ in Eq. (\ref{eq:gii})].
(ii) $G_{1 1}^{h e}$: It can be Andreev-reflected as a hole, which
transfers a Cooper pair into the stripe contributing to $I_1$ [the contribution from $G^{h e}_{1 1}$ adds to
the term $\sim \mathbbm{1}$ in Eq. (\ref{eq:gij})]. (iii) $G^{e e}_{2
1}$: It can be transmitted as an electron into lead 2 by direct charge
transfer. This process contributes positively to $I_1$ (contained in $\mathbbm{1} - R^{ee}_{11}$) and negatively to
$I_2$. (iv) $G^{h e}_{2 1}$: It can be cross-Andreev reflected as a hole
into lead 2, leaving a Cooper pair in the stripe. This process contributes
positively both to $I_1$ and $I_2$.

%This completes our discussion of the scattering approach and we next discuss
%the transport spectra.

\subsection{Transport spectra: Local vs. nonlocal current
response}\label{sec:non-local}

Our results for the local (nonlocal) conductance $G_{1 1}$ ($G_{2 1}$), are
summarized in Fig.~\ref{fig:conductance}, for both the symmetric (left) and asymmetric (right) device structure as investigated in the main part. The peak positions in the conductance spectrum closely resemble the energy spectrum of the devices [Figs. 2(b) and (e) in the main part]. The individual contributions
$G^{\tau \tau'}_{i j}$ from different transport processes [see Eq.
(\ref{eq:getaetapr})] are shown in Fig.~\ref{fig:conductance_contributions} and the stripe-length dependence is shown in Fig.~\ref{fig:conductance_length}.
%In the two figures, we use a different broadening $\eta$: in Fig.
%\ref{fig:conductance} we use $\eta = \Gamma_p / 10$ and in Fig.
%\ref{fig:conductance_contributions} \ we use $\eta = 0$ for otherwise the same
%parameters. The conductance peaks in Fig. \ref{fig:conductance_contributions}
%are therefore much sharper than in Fig. \ref{fig:conductance}. The reason why
%we introduced a finite $\eta$ in Fig. \ref{fig:conductance} is that this
%reduces the number of points needed to obtain a 2D conductance map without
%missing any sharp conductance peaks.

\Figure{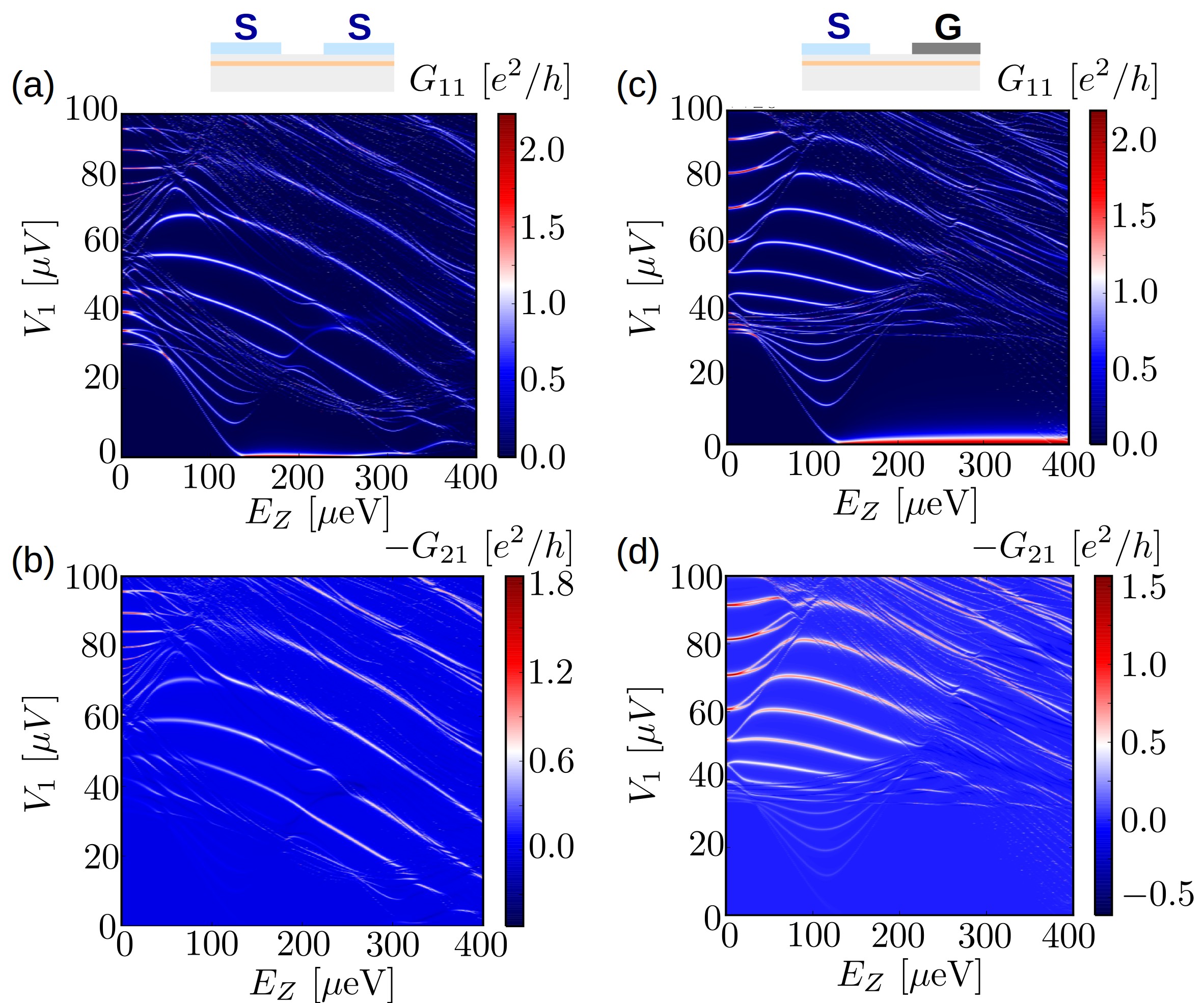}{{Identifying
  MBS from local versus nonlocal transport.} We consider a symmetric device (left) and an asymmetric device (left) as the pictograms above (a) and (c) indicate. The upper panels (a) and (c) show
  the local differential conductance $G_{1 1} = \partial I_1 / \partial
  V_1$ and in the lower panels (b) and (d) show the nonlocal differential
  conductance $G_{2 1} = \partial I_2 / \partial V_1$. 
  We assume zero
  temperature, a tunnel coupling to the probing leads of $\Gamma_p = 10~\mu$eV, $N_x =N_y =
  200$ lattice sites, and all other parameters as in Fig. 1(b) of the
  main part.\label{fig:conductance}}

\Figure{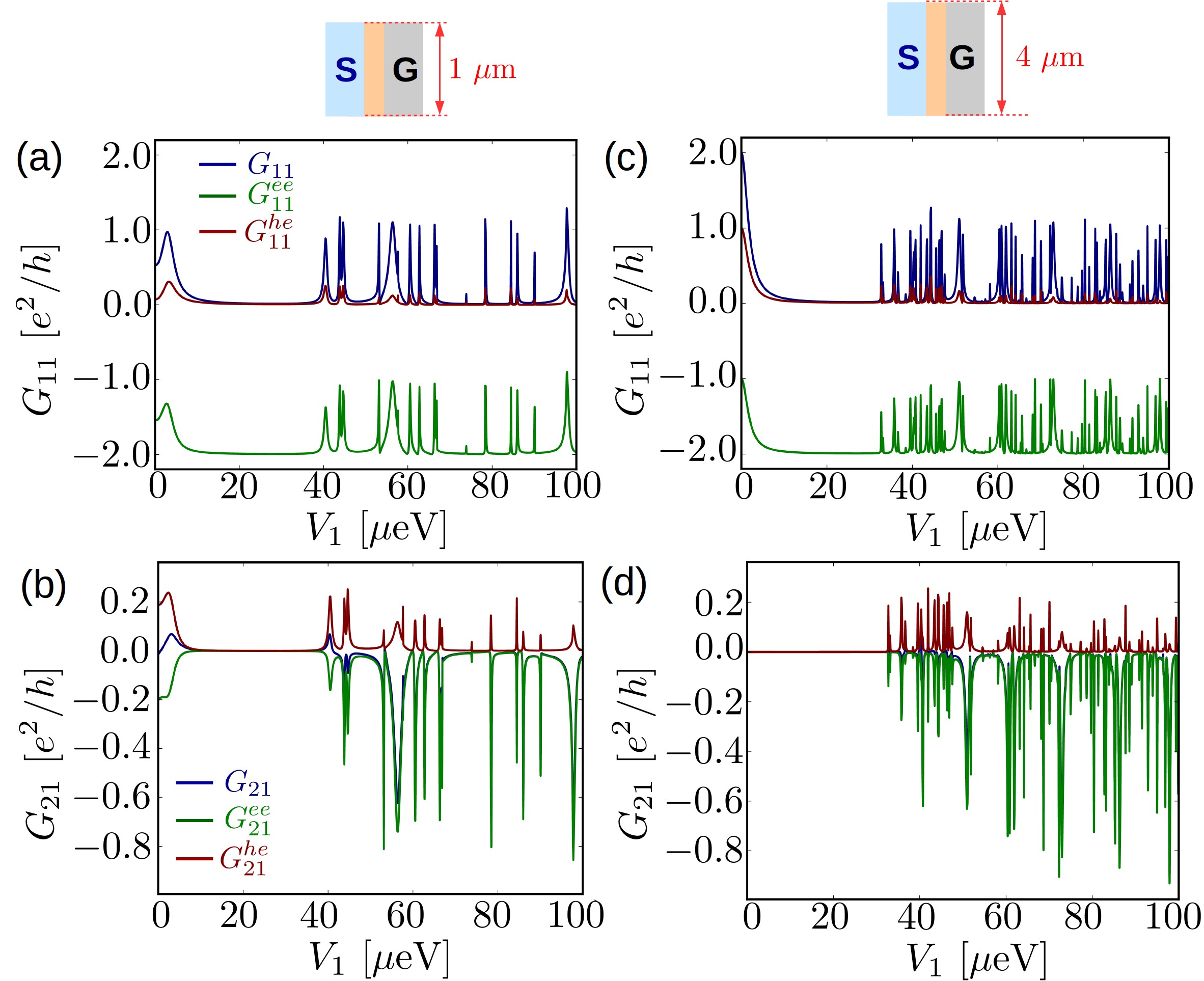}{{Transport contributions
  to the conductance.} The upper panels (a) and (c) show
  the normal reflection contribution $G_{1 1}^{e e}$ and the Andreev
  reflection contribution $G_{1 1}^{h e}$ to the local conductance $G_{11}$.
  The lower panels (b) and (d) show the direct-charge-transfer contribution
  $G_{2 1}^{e e}$ and the crossed Andreev-reflection contribution $G_{2 1}^{h
    e}$ to the nonlocal differential conductance $G_{2 1}$. All plots are for an asymmetric gated device structure ($L_R=0$) at Zeeman energy $E_Z = 200~\mu$\tmop{eV}, using 
  $L_Y = 1~\mu$m in the left panels and $L_Y = 4~\mu$m in the right panels. The right panels thus correspond to a vertical slice through right panels in \Fig{fig:conductance}. The tunnel coupling is $\Gamma_p = 10~\mu$eV, the numbers of sites are $N_x  =
  200$ and $N_y = L_Y [ \tmop{nm}] / 20$, and all other parameters are as in Fig. 1(b) of the
  main part.\label{fig:conductance_contributions}}

In our discussion, we first focus on the local response $G_{1 1} = \partial
I_1 / \partial V_1$, shown in the upper panels of Figs.~\ref{fig:conductance} and \ref{fig:conductance_contributions}. Similar to nanowires,
{\cite{ZeroBiasAnomaly1,ZeroBiasAnomaly3,ZeroBiasAnomaly4,ZeroBiasAnomaly5,ZeroBiasAnomaly6}},
we identify a zero-bias peak for a large range of Zeeman energies $E_Z$ [Fig.
  \ref{fig:conductance}(a) and (c)].
%The device is here in the topological regime for a
%large range magnetic fields because we chose a large phase bias $\varphi_2 - \varphi_1 = 3 \pi / 4$ [see Fig. 4(c) and the corresponding discussion in the
%main part]. The peak value is close to $2 e^2 / h$, where the slight reduction
%is due to choosing $\eta \neq 0$. For $\eta = 0$,
The conductance peak reaches $2 e^2 / h$ and the peak width is somewhat smaller than the tunnel coupling $\Gamma_p$ to the
leads [Fig. \ref{fig:conductance_contributions}(c)]. The reason for this is that the leads couple only to the first row of sites in the stripe but the MBS wave functions are nonzero over a larger number of sites along the stripe direction [see Fig. 1(b) in the main part].

From inspecting the individual contributions to the conductance, we can see
that normal electron reflection contributes with $G_{1 1}^{e e} \approx - 1$,
while the local Andreev reflection contributes with $G_{1 1}^{e e} \approx +
1$ [Fig. \ref{fig:conductance_contributions}(c)]. This result can be
interpreted as consequence of the spin polarization of the MBS: When the spin
of the incoming electron matches the spin of MBS, a local Andreev reflection
process happens and a Cooper pair is transferred to the stripe. However, when
the incoming electron has opposite spin direction, it must be normally
reflected. Hence, the latter species of electrons does not contribute to the
current. This explains why the MBS conductance peak rises to $2 e^2 / h$.

We further find that a zero-bias peak appears only when the stripe length exceeds
$L_Y \gtrsim 2.0 \mu$m [Fig. \ref{fig:conductance_length}(a)]. When reducing the
length, the zero-bias peak moves to finite bias [see Fig.
\ref{fig:conductance_length}(a) for $L_Y \leq 1.5 \mu$m]. The reason for this
behavior is that the MBS wave functions at the two ends start to overlap, which
leads to a coupling between the MBS and shifts their energy to a finite
value. At the same time, the local Andreev reflection is increasingly
suppressed compared to the case of zero-energy modes [Fig.
\ref{fig:conductance_contributions}(d)]. The peak value of the conductance is
thus suppressed far below $2 e^2 / h$.

\Figure{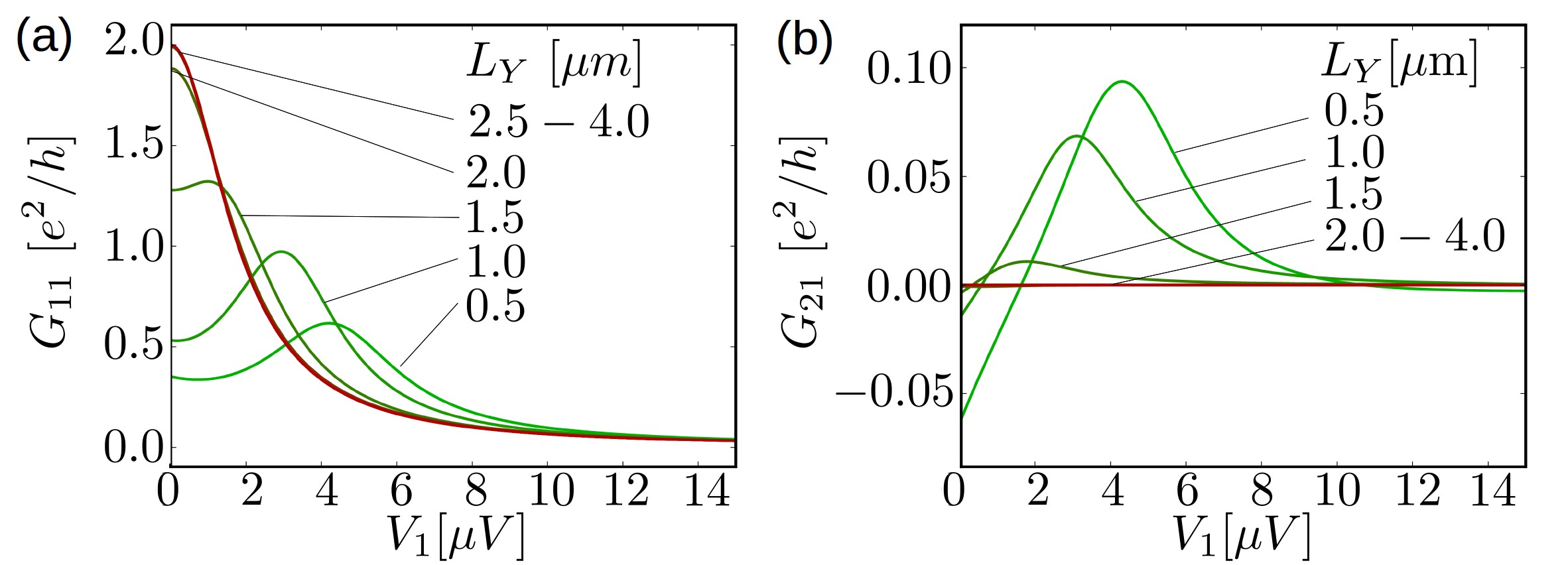}{{Spectroscopic signatures of MBS splitting depending on the stripe length.} Panel (a) shows
  the local differential conductance $G_{1 1} = \partial I_1 / \partial
  V_1$ and panel (b) shows the nonlocal differential
  conductance $G_{2 1} = \partial I_2 / \partial V_1$. We consider an asymmetric gated device structure ($L_R=0$) and vary the stripe length $L_Y$ in steps of 0.5~$\mu$m from 0.5--4.0~$\mu$m.  The Zeeman energy is $E_Z = 200~\mu$\tmop{eV}, $\Gamma_p = 10~\mu$eV, the number of lattics sites $N_x =
  200$ and $N_y = L_Y [ \tmop{nm}] / 20$, and all other parameters are as in Fig. 1(b) of the
  main part. \label{fig:conductance_length}}

We next discuss the nonlocal conductance $G_{2 1} = \partial I_2 / \partial
V_1$ depicted in the lower panels of Figs.~\ref{fig:conductance} and
\ref{fig:conductance_contributions}. Starting again with a long stripe, we see that the
zero-bias peak is clearly absent in the nonlocal response in contrast to the local response [Fig.
\ref{fig:conductance}(b) and (d)]. Moreover, the
two contributions for direct charge transfer, $G_{21}^{e e}$, and crossed
Andreev reflection, $G_{21}^{h e}$, are both individually suppressed for low
energies [Fig. \ref{fig:conductance_contributions}(d)]. In agreement with the
literature {\cite{Nilsson08}}, we verified that the ratio $G_{21}^{e e} / G^{h
e}_{21} \rightarrow - 1$ in the limit of long stripes when the particle and
hole weight of the MBS are equal (not shown). The reason for the strong
suppression is the exponentially small overlap of the localized MBS. This is
different from other subgap states, which are extended along the stripe and
thus allow for nonlocal transport [Fig. \ref{fig:conductance}(b) and (d)]. The
transport through these states is dominated by direct charge transfer [Fig.
\ref{fig:conductance_contributions}(d)].

The nonlocal conductance spectra are particularly useful to estimate the
topological gap when accounting for finite-frequency corrections to the self
energy. The topological gap obtained from the
positions of first finite-bias peak [Fig. \ref{fig:conductance_contributions}(b) and (d)] at about $35~\mu$V are close to the topological gaps $E_g$ of about $40~\mu$V obtained in the
main part without finite-frequency corrections [Fig. 3(c) and (d)]. This shows that the reduction due to
finite-frequency corrections is indeed rather small. This is also expected
because finite-frequency corrections play an important role only when energies
approach the superconducting gap $\Delta$ [see Eq. (\ref{eq:reddelta})].

Finally, for shorter stripe lengths $L_Y \lesssim 1.5~\mu$m, a low-energy peak
appears in the nonlocal response at the same energy as in the local response
[Fig. \ref{fig:conductance_length}(b)]. This is consistent with the picture of an
increased MBS wave function overlap. The net nonlocal current results from a
competition of direct charge transfer contribution $G_{21}^{e e}$ and the
contribution $G_{21}^{h e}$ from crossed Andreev reflection.
This leads to a
considerable reduction of the net current [Fig. \ref{fig:conductance_contributions}(b)]. However,
direct charge transfer dominates in particular for larger bias voltages
because the character of the finite-energy states becomes more and more
electron- and less hole-like.

In summary, our transport calculations indicate that the MBS in the Majorana
stripe could be probed experimentally similar to nanowire setups. However, we
assumed in our calculations that the transport is coherent over the
entire stripe length, which is probably hard to achieve in an experiment. It
could therefore be interesting to investigate how stable the above transport
features would be when including disorder into the calculation. Moreover, the
probes are exposed to magnetic fields and exhibit spin-orbit coupling,
which may also affect the transport features.

\section{Magnetic-field angle dependence of topological gap}

\Figure{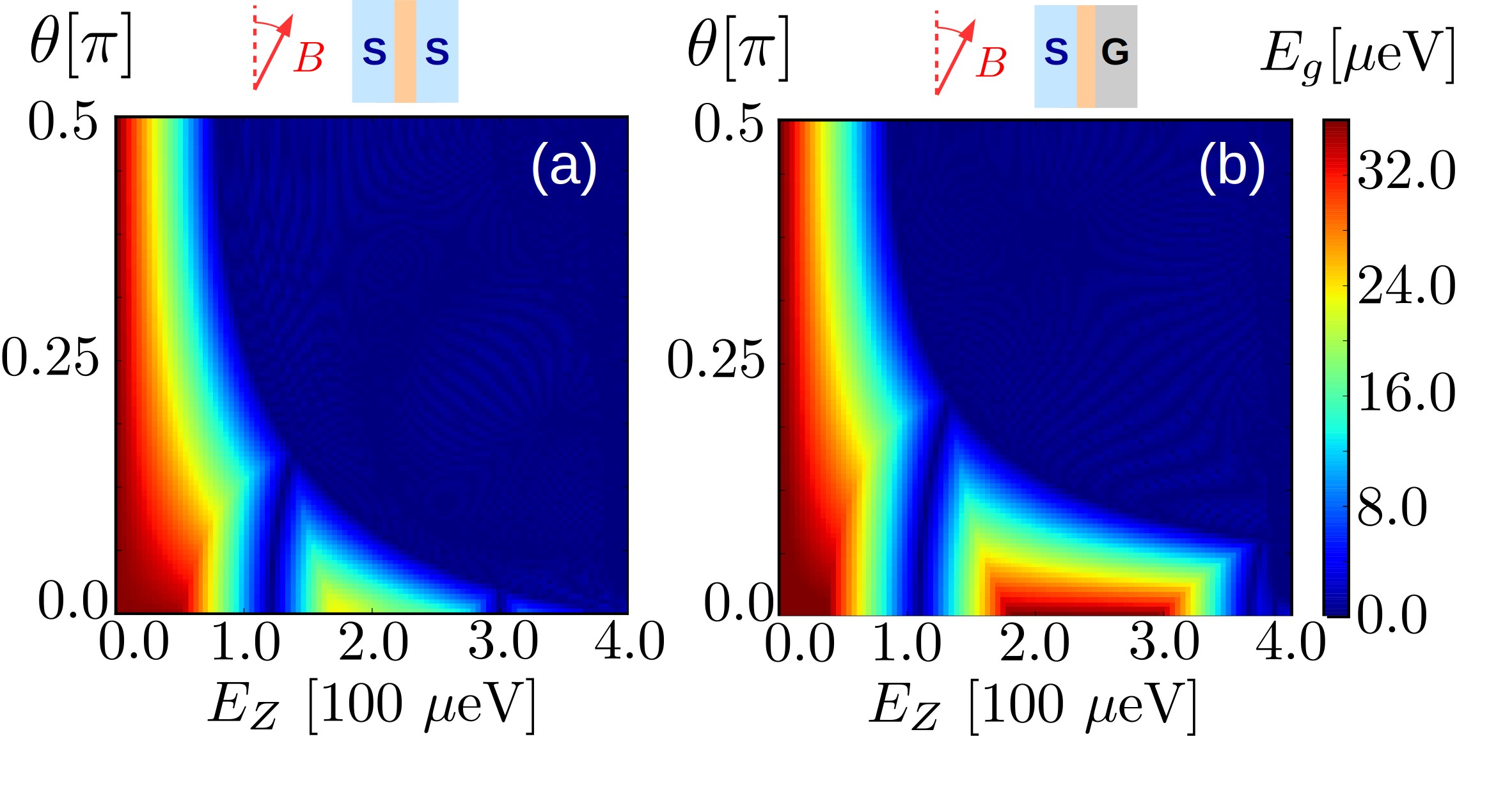}{{Magnetic-field-angle
  stability of the topological phase.} The plots show the topological gap
  $E_g = \min_{n, k_y} E_n ( k_y) \nocomma$, where $E_n ( k_y)$ are the
  eigenvalues of the 1D effective Hamiltonian (\ref{eq:heff1d}), but with the
  Zeeman term replaced by Eq. (\ref{eq:hz}). The angle dependence is shown in
  (a) for a symmetric device $( L_L = L_R)$ and in (b) for an asymmetric
  device ($L_R = 0$). We chose $N_x = 500$ and all parameters as in Fig. 1(b)
  of the main part.\label{fig:angle_dep}}

In this Section, we investigate the stability of the topological phase and
the topological gap when the magnetic field is rotated in the plane of the
2DEG. For this purpose, we replace the Zeeman term in Eq. (\ref{eq:heff1d}) by
\begin{eqnarray}
  H_Z & = & \sum_{n_x n_y \tau \sigma} \frac{E_Z}{2} [ \sin ( \theta) + i \sigma \cos (
    \theta)] | n_x n_y \tau \bar{\sigma} \rangle \langle n_x n_y \tau \sigma | . \nonumber \\
  & &  \label{eq:hz}
\end{eqnarray}
The case $\theta = 0$ corresponds to a magnetic field along the stripe, which
we investigated so far.

We find that the topological gap is suppressed when the magnetic field is
rotated away from the stripe direction [Fig. \ref{fig:angle_dep}]. For a
symmetric device structure, the topological regime breaks down for an angle of
about $\theta \approx 10^{\circ}$ close to the first phase-transition point
and for even smaller angles when the Zeeman energy approaches the second
phase-transition point [Fig. \ref{fig:angle_dep}(a)]. By contrast, the
topological regime is more stable against rotations of the magnetic-field
direction for an asymmetric device design: Here, the topological regime
persists up to $\theta \approx 20^{\circ}$. Hence, an asymmetrically gated
structure is also favorable to stabilize the MBS against a misalignment of
the magnetic field.

Irrespective of the design, we conclude, however, that a network of MBS in a
2DEG structure should contain all Majorana stripes in parallel. This is an
important restriction for the device design that we accounted for in our
device suggestions in Fig. 4 in the main part.

\putbib[cite]
%\bibliography{cite}
%\bibliographystyle{apsrev}
\end{bibunit}

\end{document}